\documentstyle[preprint,aps]{revtex}
\begin{document}
\draft
\preprint{\vbox{
\hbox{TRI-PP-93-72}
\hbox{August 1993}
}}
\title{
Can $\mu$--$e$ Conversion in Nuclei be a Good Probe for \\
Lepton-Number Violating Higgs Couplings ?
}

\author{Daniel Ng and John N. Ng}
\address{
TRIUMF, 4004 Wesbrook Mall\\
Vancouver, B.C., V6T 2A3, Canada
}
\maketitle
\thispagestyle{empty}

\begin{abstract}
Motivated by the improving sensitivity, $R$, of experiments on
$\mu~Ti \rightarrow e~Ti$ and the enhanced Higgs nucleon interaction,
we study this lepton number violating process induced by Higgs exchange. Taking
the possible sensitivity, $R \simeq 10^{-16}$, we obtain the
constraint on the Higgs-muon-electron vertex, $\kappa_{\mu e}$, to be
less than $2.4\times10^{-7}$ if the masses of the Higgs scalar and $W$
gauge boson are the same.  $\kappa_{\mu e}$ is also calculated for some models.
\end{abstract}

\newpage

In this paper we report on a study of direct muon-electron conversion in
nuclei as a probe of new physics represented by an effective $\mu-e-S$
vertex where $S$ is a neutral scalar particle.  For example, the scalar
$S$ can be the Higgs scalars in a one doublet model with an extended
fermion sector, or linear combinations of scalar particles in some extended
Higgs models such as the supersymmetric standard model.  Previous discussion
of $\mu-e$ conversion concentrated mostly on the effects of virtual
photon and $Z$--boson exchanges \cite{shanker}.  Effects of an extra
$Z$--boson has also been considered recently \cite{bernabeu}.  Scalar
and pseudoscalar effects were outlined in Ref. \cite{shanker} where the
details of the nuclear effects were emphasized.  If the
Higgs coupling to nucleon is taken to be proportional to the current
masses of the $u$-- and $d$--quarks, then the effect would be very small.
Here, we treat the scalar-nucleon-nucleon via the approach of Shifman, et.
al. \cite{shifman}, which increase the coupling strength to that of
$\frac{2}{27}m_N$ where $m_N$ is the mass of the nucleon.  This is
approximately one order of magnitude enhancement over the use of the
current quark masses. A second enhancement of the $S-N-N$ coupling can
arise in extended Higgs model where it is multiplied by ratios
of vacuum expectation values of scalars fields.
In supersymmetry, the ratio $\tan\beta \geq 10$ is certainly acceptable.
The third factor comes from the fact that scalar exchange in $\mu-e$ is
coherent
\footnote{Such is not the case for pseudoscalar and axial vector
exchanges and henceforth we shall ignore them}
over the nuclei \cite{goodman}.  In this respect, it is
similar to photon and Z exchange.

On the experimental side, we are encouraged by the on going experiment at PSI
\cite{psi} of $\mu ~ Ti \rightarrow e ~ Ti$ which will achieve a
sensitivity $R(Ti)=\Gamma({\mu Ti \rightarrow e Ti})/\Gamma({\mu Ti
\rightarrow \nu_\mu Ti})\simeq 3\times10^{-14}$ and prospects of lowering
this limit to the level of $10^{-15}-10^{-16}$ being considered at INS
Moscow \cite{melc} and TRIUMF \cite{triumf}.  This motivated us to reexamine
$\mu-e$ conversion and focus on it as a probe of the non-standard
$\mu-e-S$ vertex.  Comparison with $\mu \rightarrow e~\gamma$ and/or
$\mu \rightarrow 3e$ where they appliy are also given.

At the quark level, the effective
interaction Lagrangian induced by an exchange of a scalar $S$ is given by
\begin{equation}
{\cal L}_q(S) = \frac{G_F}{\sqrt2}~\frac{m_W^2}{m_S^2}~
        \bar e~\left[~\kappa_{\mu e}~\left(1+\gamma_5\right)~+~
   ~\kappa'_{\mu e}~\left(1-\gamma_5\right)\right]~\mu
      \sum_{\scriptstyle q=all} \frac{m_q}{m_W}~\lambda_q~\bar q q \ ,
\end{equation}
where $\kappa_{\mu e}$ and $\kappa'_{\mu e}$
are coefficient of the effective $\mu-e-S$ vertex
and $G_F$ is the Fermi coupling constant.  $m_S$ and
$m_W$ are the masses for the scalar $S$ and the standard W gauge boson
whereas $m_q$'s are the current quark masses and the sum is taken over
for all quark flavors of a given model.  For extended Higgs
models, $\lambda_q$ is not equal to unity as in the standard model.  In
particular, in the two Higgs doublet extension of the standard model
with natural flavor conservation, we have
$\lambda_{up} = \cot\beta$ and $\lambda_{down} = \tan\beta$. These
correspond to the linear combination given by
$S=-\sin\beta \sqrt2 Re \phi_1^0 + \cos\beta \sqrt2 Re \phi_2^0$, where
$\phi_1^0$ and $\phi_2^0$ are the neutral components of the Higgs
doublets that provide masses for $down$-- and $up$--type quarks
separately.  In order to see how the discussed factors enter into the
study of lepton number violation in general and the $\mu-e-S$ vertex in
particular, we first study the cases where the scalar $S$ couples to
quark like that of the standard model Higgs boson, $H$.  The effects
of Higgs mixing will be illustrated by the minimal supersymmetric model
with lepton number violation added in.

To compute the interaction at the nucleon level, we follow the procedure
suggested by Shifman et. al. \cite{shifman}.  Including the effects of
the strange and heavy quark contributions \cite{cheng,gasser},
we obtain the effective
Lagrangian for $\mu-e$ conversion in nuclei as follows,
\begin{equation}
{\cal L}_N(S) = \frac{G_F}{\sqrt2}~\frac{m_W^2}{m_S^2}~
    \bar e\left[~\kappa_{\mu e}~\left(1+\gamma_5\right)~+~
   ~\kappa'_{\mu e}~\left(1-\gamma_5\right)\right]~\mu
      ~\frac{\tilde m_N}{m_W}~\bar\Psi_N\Psi_N \ ,
\end{equation}
and
\begin{equation}
\label{mn}
{\tilde m_N}= \frac{2}{27} n_h m_N + \left(1+\frac{y}{2}\frac{m_s}{\bar
m} \right) \left(1- \frac{2}{27} n_h \right) \sigma_{\pi N} \ ,
\end{equation}
where $\Psi_N$  is the nucleon wave function.
$n_h$ is the number of heavy quarks other than $u$, $d$ and $s$.
$y$ is the strange content in the nucleon and $\sigma_{\pi N}$ is nucleon
matrix element of the $\sigma$ term in the chiral Lagrangian.
$m_N$ is the nucleon mass, $\bar m = (m_u+m_d)/2$,  and we take
${m_s}/{\bar m} \simeq 25$.  The quantity $\tilde m_N$ conveniently
expresses the heavy quark effects in Higgs-nucleon-nucleon coupling.
Its value depends on $y$ and $\sigma_{\pi N}$, where $(y,\sigma_{\pi N})
= (0,0),~(0.47,60{\rm MeV})$ and $(0.22,45{\rm MeV})$ are used in
Refs.~\cite{shifman}, \cite{cheng} and \cite{gasser} respectively.
Particularly, $\tilde m_N = 350 \rm MeV$ for
$(y,\sigma_{\pi N})=(0.22,45{\rm MeV})$ for $n_h=3$.
Hence, $\tilde m_N$ is almost two orders of magnitude
bigger than the current quark masses of $u$ and $d$.  This a a larger
enhancement factor than originally anticipated as discussed in the
introduction.

Using the standard procedure \cite{feinberg}, we obtain the conversion rate of
$\mu~N \rightarrow e~N$ as follows,
\begin{eqnarray}
\Gamma(\mu N \rightarrow e N)=
    &&\frac{G_F^2}{2}\left(\frac{\tilde m_N}{m_W}\right)^2
  \frac{\alpha^3 m_\mu^5 Z_{eff}^4}{\pi^2 Z}A^2 |F(q^2)|^2 \frac{m_W^4}{m_S^4}
\nonumber \\
     &&	\int \left[ |\kappa_{\mu e}|^2
          \frac{\left(1-{\bf s_{\mu}} \cdot {\bf \hat p_e} \right)}{2}+
            |\kappa'_{\mu e}|^2
          \frac{\left(1+{\bf s_{\mu}} \cdot {\bf \hat p_e} \right)}{2}\right]
          d \cos\theta
\end{eqnarray}
where $A$ and $Z$($Z_{eff}$) are the nucleon and (effective) atomic numbers.
$F(q^2)$ is the nucleon form factor.  $\bf s_\mu$ and $\hat p_e$ are the
muon spin and the direction of the outgoing electron.
Particularly, $Z_{eff}=17.6$ \cite
{pla} and $F(q^2=-m_{\mu}^2)=0.54$ \cite{frois} for $^{22}_{48}Ti$.
Using the muon capture rate in $Ti$, $\Gamma({\mu Ti \rightarrow \nu_\mu Ti})=
2.590\pm 10^6 sec^{-1}$ \cite{suzuki}, we obtain
\begin{equation}
\label{const}
\left(|\kappa_{\mu e}|^2+|\kappa'_{\mu e}|^2\right)^{1/2}
  \leq 2.4\times10^{-7} \left(\frac{0.5 {\rm GeV}}{\tilde m_N}\right)
   \left(\frac{R}{10^{-16}}\right)^{1/2}\frac{m_S^2}{m_W^2} \ .
\end{equation}
This is the model independent constraint on
$\kappa_{\mu e}$ and $\kappa'_{\mu e}$.
In Ref.~\cite{shanker}, the author obtained the constraint on
$\left(|\kappa_{\mu e}|^2+|\kappa'_{\mu
e}|^2\right)^{1/2}\frac{\tilde m_N}{m_W}\frac{m_W^2}{m_S^2} \leq
10^{-6}$ for sulphur.  If we take the current quark mass approach,
namely $\tilde m_N = (m_u+m_d)/2 = 5 \rm MeV$,
it yields $\left(|\kappa_{\mu e}|^2+|\kappa'_{\mu
e}|^2\right)^{1/2} \leq 1.6\times 10^{-2}$ assuming $m_S=m_W$.
Even with the improved sensitivity of two orders of magnitude,
the constraint is no better than $10^{-5}$ if
the current quark masses are used.  Obviously, the improved calculation
of Eq.~(\ref{mn}) gives a much better limit as evident from
Eq.~(\ref{const}).

Encouraged by the enhancement of the Higgs nucleon interaction, we
study three examples to see when this lepton number violating Higgs
interaction be important for the muon--electron conversion in nuclei.

{\parindent=0pt 1. \underline {Exotic Leptons}}

In the standard model, the Yukawa interactions of Higgs $H$ and leptons
are flavor diagonal, and is given by
\begin{equation}
-\frac{g}{2m_W}~H~\left[m_e \bar e e + m_\mu \bar \mu \mu + m_\tau \bar \tau
\tau \right] \ .
\end{equation}
When we include exotic leptons which mix with the ordinary leptons,
there will be lepton flavor changing Higgs interactions.  In the lowest
order, the coefficient of the $\mu-e-H$ vertex is given by
\begin{equation}
\kappa_{\mu e} \simeq \frac{1}{m_W}\left( m_e U_{e\mu} + m_\mu U_{\mu e}
+ m_\tau U_{\tau e}U_{\tau \mu} \right)\ ,
\end{equation}
where $U_{\alpha\beta}$ is the mixing in the charged lepton sector
induced by exotic leptons.
Using the constraint in Eq.~(\ref{const}), we obtain $U_{e\mu} \leq
0.04$, $U_{\mu e} \leq 2\times 10^{-4}$ and $U_{\tau e}U_{\tau \mu}\leq
1\times 10^{-5}$.  From $\mu \rightarrow 3 e$
and $\tau \rightarrow 3 \ell $
\footnote{ Note that when there are tree level lepton flavor changing
interactions, the processes $\mu \rightarrow e~\gamma$ induced at
one-loop level are less important than $\mu \rightarrow 3 e$
and $\tau \rightarrow 3 \ell$.}
by tree level exchanges of $Z$ gauge boson,
the constraints are $U_{\mu e} \leq
3\times10^{-6}$ and $U_{\tau e}U_{\tau \mu}\leq 1.6\times10^{-3}$
\cite{london}.  Hence, the future $\mu-e$ conversion experiments can
improve the constraint on the $\tau-\mu$ and $\tau-e$ mixings by two
orders.

In the following two examples, we consider models with lepton flavor
conservation at tree level.  Hence both the $\mu-e-S$ and $\mu-e-\gamma$
vertices are induced at one-loop level.

{\parindent=0pt 2. \underline {4th Generation Standard Model}}

When the standard model is extended to include the 4th generation, a
right-handed neutrino is necessary to provide the mass for the 4th
neutrino which must be heavier than $45 {\rm GeV}$ from the LEP experiments,
leading to three massless and one massive neutrinos.  In this model, the
scalar $S$ is the standard model Higgs, $H$.  Since the
vertex $\mu-e-H$ is induced
by the $V-A$ current, hence $\kappa'_{\mu e} =0$; whereas $\kappa_{\mu
e}$ \cite{grzadkowski} is given by
\begin{equation}
\kappa_{\mu e} = \frac{g^2}{16\pi^2}U_{\mu 4}U_{e 4}\frac{m_\mu}{m_W}
   \left[ \frac{3}{4}x+\frac{m_H^2}{m_W^2}\left( \frac{3x-x^2}{8(x-1)^2}
       +\frac{x^3-2x^2}{4(x-1)^3}\ln x \right) \right] \ ,
\end{equation}
where $x=m_{\nu_4}^2/m_w^2$ and $U_{\alpha\beta}$ is the CKM matrix in
the lepton sector of four flavors.   For the decay
$\mu \rightarrow e~\gamma$, the decay rate is given by
\begin{equation}
\Gamma(\mu \rightarrow e \gamma)=\frac{\alpha^3m_\mu}{64\pi\sin^4\theta_W}
\frac{m_\mu^2}{m_W^2}|U_{\mu 4}U_{e 4}I(x)|^2
\end{equation}
and
\begin{equation}
I(x)=\frac{-x+5x^2+2x^3}{4(1-x)^3}+\frac{3x^3}{2(1-x)^4}\ln x \ .
\end{equation}
where $U_{\mu 4}U_{e 4} \leq 3\times10^{-3}$ for
$m_{\nu_4} \geq 45 \rm GeV$ are obtained \cite{acker}.  However,
$\kappa_{\mu e}$ is suppressed by ${m_\mu}/{m_W}$.  Unless $m_{\nu_4}$
is greater than $2$ TeV,  the processes
$\mu \rightarrow e~\gamma$ is more important to probe
the lepton number violation mechanism.

{\parindent=0pt 3. \underline{Minimal Supersymmetric Standard Model}}

In the minimal supersymmetric extension of the standard model (MSSM), the
lepton number processes can be induced through the slepton mixing.  In
analogy to the Yukawa interactions, there exist soft-breaking terms,
$A Re\phi^0_1 {\tilde e_L}^\ast {\tilde \mu_R} + h.c.$.  Therefore the
mass matrix in the basis $\{{\tilde e_L},{\tilde \mu_R}\}$ is given by
\begin{equation}
\pmatrix { \tilde m_e^2 & A v_1 \cr A v_1 & \tilde m_{\mu}^2 \cr} \ .
\end{equation}
yielding the decay rate for $\mu \rightarrow e~\gamma$ to be
\begin{equation}
\Gamma(\mu \rightarrow e \gamma)=\frac{\alpha^3 m_\mu}{256\pi^2\sin^4\theta_W}
\sin^22\theta |\Gamma_{\mu e \gamma}|^2 \ ,
\end{equation}
where
\begin{equation}
\Gamma_{\mu e \gamma}=\sum_{\scriptstyle i=1}^{\scriptstyle 4} \tan\theta_W
N_{1i}\left(N_{2i}+N_{1i}\tan\theta_W\right) \frac{m_\mu}{m_{\chi_i}}
\left[ x_i F(x_i) - y_i F(y_i) \right] \ ,
\end{equation}
with
\begin{equation}
F(x)=-\frac{1+x}{2(x-1)^2}+\frac{x\ln{x}}{(x-1)^3} \ .
\end{equation}
$\sin\theta$ and $N_{ij}$ are the scalar and neutralino mixing parameters.
$x_i=m_{\chi_i}^2/m_1^2$ and $y_i=m_{\chi_i}^2/m_2^2$,
where $m_{\chi_i}$ and $m_{1,2}$ are the neutralino and slepton masses.

The $\mu-e$ conversion in nuclei is induced \cite{comment} by the
the vertex, $\sqrt{2}Re\phi^0_1 {\tilde e_L}^\ast {\tilde \mu_R}$,
leading to the coefficient of the effective vertex
$\mu-e-\sqrt{2}Re\phi^0_1$,
\begin{eqnarray}
\kappa_{\mu e} = &&\frac{g}{32\pi^2} \sum_{\scriptstyle i=1}^{\scriptstyle 4}
 \tan\theta_W N_{1i}\left(N_{2i}+N_{1i}\tan\theta_W\right)
  \frac{A}{\sqrt{2} m_{\chi_i}}  \nonumber \\
 && \left\{ \sin^22\theta \left[ x_i G(x_i)+ y_i G(y_i) \right]
    + 2\cos^22\theta \left[x_i H(x_i,m_2^2/m_1^2)\right] \right\}\ ,
\end{eqnarray}
where
\begin{eqnarray}
G(x)=&&\frac{1}{1-x}+\frac{x\ln{x}}{(1-x)^2} \ ,\\
H(x,r)=&&\frac{x\ln{x}}{(x-1)(x-r)}+\frac{r\ln{r}}{(r-1)(r-x)} \ ,
\end{eqnarray}
and
\begin{equation}
\label{nucmass}
\frac{\tilde m_N}{m_S^2} =
\left[\cos\beta\frac{m_A^2+m_Z^2\sin^2{2\beta}}{m_A^2m_Z^2\cos^2{2\beta}}
   -\sin\beta \frac{\tan{2\beta}}{m_A^2} \right] \tilde m_{N_1}
+\left[\cos\beta\frac{\tan{2\beta}}{m_A^2}
  -\sin\beta\frac{1}{m_A^2}\right] \tilde m_{N_2}\ ,
\end{equation}
where the effective nucleon mass induced by interacting with
$\cos\beta\sqrt{2}Re\phi^0_1+\sin\beta\sqrt{2}Re\phi^0_2$ and
$-\sin\beta\sqrt{2}Re\phi^0_1+\cos\beta\sqrt{2}Re\phi^0_2$ are given by
\begin{eqnarray}
\tilde m_{N_1}&=&\frac{2}{9}m_N+\frac{7}{9}
  \left(1+\frac{y}{2}\frac{m_s}{\bar m}\right)\sigma_{\pi N} \ ,\\
m_{N_2}&=& -\frac{2}{27}\left(\tan\beta-2\cot\beta\right)m_N
  -\left(\frac{4}{27}\cot\beta+\frac{25}{27}\tan\beta\right)
 \left(1+\frac{y}{2}\frac{m_s}{\bar m}\right)\sigma_{\pi N}\ .
\end{eqnarray}
The square brackets in Eq.~(\ref{nucmass}) are the effective Higgs
propagators.

In table \ref{table}, we tabulate the branching ratio for
$\mu \rightarrow e~\gamma$ and $\mu~Ti \rightarrow e~Ti$ in MSSM for
different values of $A$ and $tan\beta$.  We take $(y,\sigma_{\pi N}) =
(0.22,45\rm GeV)$.  For a large $\tan\beta$,
$v_1=\sqrt{v_1^2+v_2^2}\cos\beta$ is small.
thus the Higgs interaction would be at least as important as
$\mu \rightarrow e~\gamma$.  Particularly, for
 $A=500{\rm GeV}$, $\tan\beta=50$ and an intermediate mass scalar
$m_A=250\rm GeV$,
the process $\mu~Ti \rightarrow e~Ti$ is about $4$ times below
the present experimental
limit \cite{ahmad}; whereas the branching ratio for the process $\mu
\rightarrow e~\gamma$ is $20$ times below the present experimental
values \cite{bolton}.  This is especially relevant when the
sensitivity of the former is improved by two orders of magnitude;
whereas we do not foresee a similar improvement in the $\mu \rightarrow
e~\gamma$ measurement.

In conclusion, we have considered the $\mu-e$ conversion in nuclei
induced by Higgs exchange for three popular models.  This process would
be negligible if the Higgs nucleon coupling is taken to be proportional
to the current quark masses.
Here,  we have shown how the Higgs nucleon interaction is
enhanced by using the approach first employed by Shifman, et. al., and
this yields $\kappa_{\mu e} \leq 2.4\times10^{-7}$.
$\kappa_{\mu e}$ in a model of 4th generation lepton is small because it is
suppressed by the muon mass.  On the other hand, with the existence of
the soft breaking terms in the MSSM, the Higgs induced $\mu-e$
conversion is at least as important as $\mu \rightarrow e~\gamma$.
The process will be more important for a larger $\tan\beta$ as the rate
increases as the square of this parameter.  Furthermore, we have shown
that $\mu-e$ conversion can be a sensitive probe to scalar particles in
the mass range of hundreds of GeV even when the lepton-number violation
is an one-loop effect.  The minimal supersymmetric standard model is used
as an illustrative example.

This work was supported in part by the Natural Science and Engineering
Council of Canada.

\newpage
\vspace{1cm}
\begin{table}
\caption{The branching ratio for $\mu \rightarrow e~\gamma$ and $\mu~Ti
\rightarrow e~Ti$ in MSSM.  we take $\tan\beta=10(50)$, $m_A = 250 \rm GeV$
and $\tilde m_{e,\mu} = 5 \rm TeV$. For the gaugino masses, we take
$2M_1=M_2=\mu=250 \rm GeV$.}
\vspace{1cm}
\begin{tabular}{|c|c|c|} \hline
$A$(GeV) & $\mu \rightarrow e \gamma~^{\rm a}$
   & $R(\mu Ti\rightarrow e Ti)~^{\rm b}$ \\ \hline
$500$ & $4.7(0.2)\times10^{-11}$ & $0.05(1.0)\times10^{-12}$   \\
$250$ & $11(0.5)\times10^{-12}$ & $0.13(2.5)\times10^{-13}$   \\
$50$ & $46(1.9)\times10^{-14}$ & $0.05(1.0)\times10^{-14}$   \\
\end{tabular}
\label{table}
\tablenotetext[1]{the present experiment limit \cite{bolton} for the branching
ratio is $\leq 4.9\times10^{-11}$}
\tablenotetext[2]{the present experiment limit \cite{ahmad} for the process
relative to the muon capture in $Ti$ is $\leq 4.6\times10^{-12}$}
\end{table}


\begin{references}
\bibitem{comment} We neglect the effective $\mu-e-\sqrt{2}Re\phi^0_{1,2}$
vertices induced by attaching $\sqrt{2}Re\phi^0_{1,2}$ to the neutralino line
because they are negligible if the constraint from
$\mu \rightarrow e~\gamma$ is imposed.
\end{references}
\end{document}